\documentclass[preprint,12pt]{article}

\usepackage{amssymb}
\usepackage{amsmath}
\usepackage{bm}
\usepackage{natbib}

\title{A Matrix-Variate Log-Normal Model for  Covariance Matrices} 
\author{Edoardo Otranto\\
	Department of Social Sciences and Economics\\
	Sapienza University of Rome\\
	Piazzale Aldo Moro, 5; 00185 Rome\\
	e-mail: edoardo.otranto@uniroma1.it}

\begin{document}

\maketitle

\begin{abstract}
We propose a modeling framework for time-varying covariance matrices based on the assumption that the logarithm of a realized covariance matrix follows a matrix-variate oNrmal distribution. By operating in the space of symmetric matrices, the approach guarantees positive definiteness without imposing parameter constraints beyond stationarity. The conditional mean of the logarithmic covariance matrix is specified through a BEKK-type structure that can be rewritten as a diagonal vector representation, yielding a parsimonious specification that mitigates the curse of dimensionality. Estimation is performed by maximum likelihood exploiting properties of matrix-variate Normal distributions and expressing the scale parameter matrix as a function of the location matrix. The covariance matrix is recovered via the matrix exponential. Since this transformation induces an upward bias, an approximate, time-specific bias correction based on a second-order Taylor expansion is proposed. The framework is flexible and applicable to a wide class of problems involving symmetric positive definite matrices.
\end{abstract}



{\bf keywords:} Positive definite matrix, curse of dimensionality, vectorization, delta method, bias correction, BEKK model




\section{Introduction}
\label{sec:intro}
Modeling time-varying covariance matrices is a topic of growing interest in the statistical literature, particularly in financial applications, where the dynamics of conditional variances and covariances help identify and forecast periods of quiet and turmoil in financial markets, as well as  comovements and spillover effects \citep[see, for example,][]{Bauwens_Otranto:2025}. In finance, interest in modeling covariance matrices has further increased following the introduction of accurate measures of realized covariance based on ultra-high-frequency data \citep{ABCD2013}.

There are two main challenges in modeling covariance matrices \citep{BLRsurvey06}. First, the model output must be a positive definite matrix. Second, the number of estimated coefficients typically increases rapidly with the number of variables involved, leading the so-called  {\it curse of dimensionality}.

The first issue is generally addressed through parameterizations that impose various constraints on the coefficients, often resulting in a loss of flexibility. A simple and effective alternative is to apply the logarithmic transformation, which maps the space of positive definite matrices into the space of real symmetric matrices \citep{Chiu_Leonard_Tsui:1996}, thereby bypassing the need for parameter constraints other than those required for stationarity. The second issue is often tackled by imposing simple model structures, such as scalar or diagonal parameterizations. More recent solutions include the use of the Hadamard exponential parameterization \citep{Bauwens_Otranto:2025} or the identification of parameter groups based on clustering techniques \citep{otr10} which ensure both  parsimony and flexibility.

In this paper, we propose a new theoretical framework based on the assumption that the logarithm of the covariance matrix follows a matrix-variate Normal distribution. This framework is used to develop a new BEKK-type model \citep{Engle_Kroner95}, which can be easily estimated by rewriting it as a diagonal VEC model \citep{BollerslevEngleWooldridge88} and exploiting the properties of the matrix-variate Normal distribution. By reparameterizing the scale parameter matrix as a function of the location parameter matrix, we overcome the curse of dimensionality problem. Moreover, the bias induced by the exponential transformation is approximately corrected through a procedure inspired by the {\it delta method} \citep{Verhoef:2012}.  

The paper is organized as follows. Section \ref{sec:model} presents the proposed modeling framework, with two subsections devoted to the estimation algorithm and the bias correction procedure.  Section \ref{sec:concl}  concludes with some final remarks.

\section{The Proposed Model}\label{sec:model}
Let $\bm{C}_t$ be an $n \times n$ realized covariance matrix computed at time $t$ ($t=1,\dots,T$). Consider its singular value decomposition: 
\[
\bm{C}_t=\bm{V}_t \bm{\Lambda}_t \bm{V}_t'
\]
where $\bm{\Lambda}_t$ is a diagonal matrix containing the eigenvalues of $\bm{C}_t$ and $\bm{V}_t$ is the matrix of  the corresponding orthonormal eigenvectors. The logarithmic transformation of ${\bm C}_t$ is defined as:
\[
{\bm \Gamma}_t=\bm{V}_t log(\bm{\Lambda}_t) \bm{V}_t'
\]
where $log(\bm{\Lambda}_t)$ is a diagonal matrix whose entries are the logarithms of the eigenvalues.

For each $t$, ${\bm \Gamma}_t$ is assumed to follow a matrix-variate Normal distribution \citep[see, for example,][]{Gupta_Nagar:2000} with parameters ($\bm{M}_t,\bm{U}_t,\bm{U}_t$), where $\bm{M}_t$ is the location parameter matrix, which coincides with the expected value, and $\bm{U}_t$ is the scale parameter matrix, which is the same for both rows and columns of ${\bm \Gamma}_t$.\footnote{In general, for a $m \times n$ matrix-variate Normal, the scale parameter matrix is different for rows and columns. In this case, it is the same given the simmetry of the covariance matrix.} 

Our objective is to model $\bm{M}_t$, using one of the most popular multivariate specifications for conditional covariance matrices, namely the BEKK model \citep{Engle_Kroner95}, adapted as follows:
\begin{equation}
\bm{M}_t=(\bm{I}_{n}-\bm{A}-\bm{B})\odot \overline{{\bm \Gamma}}+\bm{A} \odot \bm{\Gamma}_{t-1}+\bm{B} \odot \bm{M}_{t-1} \label{log-bekk}
\end{equation}
where $\bm{I}_{n}$ is the $n \times n$ identity matrix, $\overline{{\bm \Gamma}}$ denotes the sample mean of the $\bm{\Gamma}_{t}$ matrices, $\bm{A}$ and $\bm{B}$ are matrices of coefficients, and $\odot$ indicates the Hadamard (element-wise) product. Unlike the standard BEKK model, no constraints are required to ensure positive definiteness of the output matrix, as this property is automatically guaranteed by construction.
The only constraints imposed on $\bm{A}$ and $\bm{B}$ are those ensuring stationarity and invertibility of the process. 
Moreover, the unconditional mean of $\bm{M}_t$ is equal to $\overline{{\bm \Gamma}}$.

\subsection{Estimation}\label{sec:est}
By stacking the columns of ${\bm \Gamma}_t$ into an $n\times n$ dimensional vector and exploiting the relationship between the matrix-variate Normal and the multivariate Normal distributions \citep{Gupta_Nagar:2000}, we obtain:
\[
vec({\bm \Gamma}_t) \sim N_{n^2}\left(vec(\bm{M}_t),\bm{U}_t \otimes \bm{U}_t \right) 
\]

Let $\bm{\gamma_t}=vech({\bm \Gamma}_t)$ denote the half-vectorization of the lower triangular part of ${\bm \Gamma}_t$, and let $\bm{\mu}_t=vech({\bm M}_t)$. Denote  by  $\bm{U}_t^* $ the corresponding elements of $\bm{U}_t \otimes \bm{U}_t$. The model can then be written as a  diagonal VEC model of dimension $n^*=n(n+1)/2$ \citep{BollerslevEngleWooldridge88}:
\begin{equation}
    \bm{\mu}_t=(\bm{I}_{n^*}-\bm{A}^*-\bm{B}^*)\overline{{\bm \gamma}^*}+\bm{A}^* \bm{\gamma}_{t-1}^*+\bm{B}^* \bm{\mu}_{t-1} \label{log-vec}
\end{equation}
where $\overline{{\bm \gamma}^*}$ is the sample mean of the $\bm{\gamma}_{t}^*$ vectors, and $\bm{A}^*$ and $\bm{B}^*$ are diagonal matrices of coefficients. Imposing the stationarity constraints on each pair $(a^*_i,b^*_i)$, corresponding to the diagonal elements of $\bm{A}^*$ and $\bm{B}^*$, ensures that the unconditional mean of $\bm{\mu}_t$ is $\overline{{\bm \gamma}^*}$.

Maximum Likelihood estimation is performed by maximizing the joint density of $T$ multivariate Normal random vectors of dimension $n^*$. For  large $n$, estimating the covariance matrix  $\bm{U}_t^*$, would entail a dramatic increase in the number of parameters, thereby reintroducing the curse of dimensionality. 

For a symmetric matrix-variate Normal random variable, the following conditional moments hold:\footnote{The subfix $t-1$ indicates the conditioning on the information set until time $t-1$.}
\begin{equation}
\begin{array}{l}
E_{t-1}({\bm \Gamma}_t)=\bm{M}_t \\
Var_{t-1}({\bm \Gamma}_t)=\bm{U}_t trace(\bm{U}_t)
\end{array} \label{eq:mom_n}
\end{equation}
Let $\hat{\bm{M}_t}$ denote the estimate of $\bm{M}_t$ obtained by appropriately rearranging  the estimates of $\bm{\mu}_t$ into an $n \times n$ matrix. A consistent estimator of the covariance matrix of ${\bm \Gamma}_t$ is given by:
\begin{equation}
\hat{\bm{\Sigma}}_t=(1/T)\sum ({\bm \Gamma}_t-\hat{\bm{M}_t})({\bm \Gamma}_t-\hat{\bm{M}_t})' \label{eq:sigmahat}
\end{equation}
From the above expression, the matrix $\bm{U}_t$ can be estimated as:
\begin{equation}
\hat{{\bm U}_t}=\frac{\hat{\bm{\Sigma}}_t}{\sqrt{trace(\hat{\bm{\Sigma}}_t) }} \label{eq:uhat}
\end{equation}
By replacing $\bm{U}_t^*$ with the corresponding elements of $\hat{{\bm U}_t}$, the likelihood function depends only on the parameters involved in the VEC representation (\ref{log-vec}), and its maximization yields $\hat{\bm{M}_t}$. This could be achieved through an iterative procedure in which, at each step, for a fixed value of $\bm{U}_t^*$, the likelihood is maximized with respect to the unknown parameters in (\ref{log-vec}). Given these parameter estimates, an updated $\bm{U}_t^*$  is then obtained from (\ref{eq:sigmahat})-(\ref{eq:uhat}), and the likelihood is maximized conditionally on this value until convergence. The initial value of  $\hat{\bm{\Sigma}}_t$ can be set equal to the sample covariance matrix of the series ${\bm \Gamma}_t$.

The estimated covariance matrix is then recovered as:
\begin{equation}
\hat{\bm{C}_t}=exp(\hat{\bm{M}_t})=\bm{W}_t \exp(\bm{L}_t) \bm{W}_t^{-1} \label{bia_est}
\end{equation}
where $\bm{L}_t$ is the diagonal matrix of eigenvalues of $\hat{\bm{M}_t}$ and $\bm{W}_t$ the matrix of the corresponding eigenvectors.
The resulting matrix is positive definite by construction, since $\hat{\bm{M}_t}$ is symmetric.

\subsection{Bias Correction}\label{sec:bias}
 The estimator   (\ref{bia_est}) is  upward biased due to the convexity of the exponential function, as implied by the Jensen's inequality. In a related framework, \citet{Bauer_Vorkink_2011}  propose a correction based on matching realized and estimated standard deviations while leaving correlations unchanged, applying the same correction at all time points. In this paper, we also focus on correcting the variances, but we allow for time-specific adjustments based on an approximation of the bias obtained through a second-order Taylor expansion of the exponential function  (\textit{delta method}). 

Given the diagonal structure of the VEC representation (\ref{log-vec}), consider the individual equations for the $n$ realized variances ($i=1,\dots,n$):
\begin{equation}
    \hat{\mu}_{i,t}=(1-\hat{\alpha}_i^*-\hat{\beta}_i^*)\bar{\gamma}_i^*+\hat{\alpha}_i^* {\gamma}_{i,t-1}^*+\hat{\beta}_i^* \hat{\mu}_{i,t-1} \label{single_eq}
\end{equation}

Let $m_{i,t}$ denote the estimator of ${\mu}_{i,t}$. A second-order Taylor expansion of $\exp(m_{i,t})$ around $\hat{\mu}_{i,t}$ yields:
\begin{equation}
    \exp(m_{i,t})\approx\exp(\hat{\mu}_{i,t})+\exp(\hat{\mu}_{i,t})(m_{i,t}-\hat{\mu}_{i,t})+ \frac{1}{2}\exp(\hat{\mu}_{i,t})(m_{i,t}-\hat{\mu}_{i,t})^2
\end{equation}
Taking conditional expectations, we obtain
\begin{equation}
    E_{t-1}(\exp(m_{i,t}))\approx \exp(\hat{\mu}_{i,t})(1+\frac{1}{2}Var_{t-1}(m_{i,t})) \label{exp_appr}
\end{equation}
which shows that $\exp(\hat{\mu}_{i,t})$ is a biased estimate of the corresponding elements of $\bm{C}_t$. The right-hand side provides an approximately unbiased estimator.

 The conditional variance  $Var_{t-1}(\hat{\mu}_{i,t})$ can be obtained using the law of total variance:
\begin{equation}
 {Var_{t-1}(m_{i,t})}=Var(m_{i,t})-E(Var_{t-1}(m_{i,t})) \label{ltv}
\end{equation}
The first term of the right-hand side of (\ref{ltv}) can be estimated using the sample variance of the series $\hat{\mu}_{i,t}$. The second term can be computed from the estimated model parameters in (\ref{single_eq}) as:
\begin{equation}
\begin{array}{ll}
\widehat{E(Var_{t-1}(m_{i,t})})=&(\gamma_{i,t-1}^*-\bar{\gamma}_i^*)^2 Var(\hat{\alpha}_i^*)+(\hat{\mu}_{i,t-1}-\bar{\gamma}_i^*)^2 Var(\hat{\beta}_i^*)\\
&+2(\gamma_{i,t-1}^*-\bar{\gamma}_i^*)(\hat{\mu}_{i,t-1}-\bar{\gamma}_i^*)Cov(\hat{\alpha}_i^*,\hat{\beta}_i^*)
\end{array}
\end{equation}

Consider the decomposition of the estimated covariance matrix:
\begin{equation}
\hat{\bm{C}_t}=\hat{\bm{D}_t}\hat{\bm{R}_t}\hat{\bm{D}_t}
\end{equation}
where $\hat{\bm{D}_t}$ is a diagonal matrix with entries given by the estimated standard deviations $[\exp(\hat{\mu}_{i,t-1})]^{1/2}$, and $\hat{\bm{R}_t}$ is the estimated correlation matrix. Bias correction is implemented by replacing $\exp[\hat{\mu}_{i,t-1}]$ in $\hat{\bm{D}_t}$ with the approximation in (\ref{exp_appr}).

\section{Some Remarks}\label{sec:concl}
We propose a model for realized covariances that circumvents the need for parameter constraints by modeling the logarithms of symmetric matrices under the assumption of a matrix-variate Normal distribution. Realized covariances are employed because they represent one of the most accurate and widely used measures of variances and covariances in financial time series. However, the proposed approach can be applied to any alternative measure of covariance matrices, not necessarily in financial applications, or more generally to any model for symmetric positive-definite matrices.

We show how estimation can be simplified by exploiting a vectorized representation of the model, thereby reducing the number of parameters to be estimated through an explicit expression of the scale parameter matrix of the Normal distribution as a function of the location parameter matrix. When the number of variables $n$ is large, the number of coefficients in the parameter matrices $\bm{A}$ and $\bm{B}$ may still be substantial ($n(n+1)/2$ in each matrix). In such cases, a feasible specification may be obtained by further reducing the number of parameters using clustering techniques, such as those proposed for correlation models in \citet{otr10}. 

Finally, we introduce a bias correction for the covariance estimator that is time specific.

Future research will focus on implementing the proposed theoretical framework in empirical applications and assessing the accuracy of the approximations induced by reparameterization and bias correction.

\section*{Acknowledgements}
This work was supported by the
Italian PRIN 2022 grant “Methodological and computational issues in large-scale time
series models for economics and finance” (20223725WE).

  





\end{document}